# Monolayer superconductivity and tunable topological electronic structure at the Fe(Te,Se)/Bi$_2$Te$_3$ interface

Robert G. Moore[1,*], Tyler Smith[1], Xiong Yao[2], Yun-Yi Pai[1], Michael Chilcote[1], Hu Miao[1], Satoshi Okamoto[1], Seongshik Oh[2], Matthew Brahlek[1]

[1] Materials Sciences and Technology Division, Oak Ridge National Laboratory, Oak Ridge, TN 37831, USA

[2] Department of Physics and Astronomy, Rutgers, the State University of New Jersey, Piscataway, NJ 08854, USA

*Email: moorerg@ornl.gov

## Abstract

The interface between two-dimensional topological Dirac states and an *s*-wave superconductor is expected to support Majorana bound states that can be used for quantum computing applications. Realizing these novel states of matter and their applications requires control over superconductivity and spin-orbit coupling to achieve spin-momentum locked topological surface states which are simultaneously superconducting. While signatures of Majorana bound states have been observed in the magnetic vortex cores of bulk FeTe$_{0.55}$Se$_{0.45}$, inhomogeneity and disorder from doping makes these signatures unclear and inconsistent between vortices. Here we report superconductivity in monolayer FeTe$_{1-y}$Se$_y$ (Fe(Te,Se)) grown on Bi$_2$Te$_3$ by molecular beam epitaxy. Spin and angle resolved photoemission spectroscopy directly resolve the interfacial spin and electronic structure of Fe(Te,Se)/Bi$_2$Te$_3$ heterostructures. We find that for *y* = 0.25 the Fe(Te,Se) electronic structure overlaps with the topological Bi$_2$Te$_3$ interfacial states which disrupts the desired spin-momentum locking. In contrast, for *y* = 0.1 a smaller Fe(Te,Se) Fermi surface allows for clear spin-momentum locking observed in the topological states. Hence, we demonstrate the Fe(Te,Se)/Bi$_2$Te$_3$ system is a highly tunable platform for realizing

Majorana bound states where reduced doping can improve characteristics important for Majorana interrogation and potential applications.

**Introduction**

The interplay between topology and electronic correlations can fractionalize electronic degrees of freedom into exotic quasiparticles as Majorana bound states (MBS) that demonstrate non-Abelian statistics which can be used for topological quantum computation applications.[1,2] These MBS can be realized on the boundary of a superconductor with an odd-parity, e.g. spin-triplet or $p$-wave, pairing state.[3,4] However, only a few candidate $p$-wave superconductors exist, such as $Cu_xBi_2Se_3$ and $UTe_2$, which tend to have low $T_c$s and are sensitive to disorder.[5-11] Besides intrinsic odd-parity superconductors, there are several avenues to create MBS via topological superconductivity from proximity induced effects.[4] Prominent examples include superconducting materials with topological surface states and the interface between a $s$-wave superconductor and a topological insulator, where the superconducting proximity effect induces and effective $p$-wave pairing in the topological interfacial states.[4,12] Among this class of candidate topological superconductors, the Fe(Te,Se) system has attracted considerable attention.[13-21]

The Fe(Te,Se) systems has spin-helical Dirac surface states due to the topological band inversion between the $p_z$ and $d_{xy}/d_{yz}$ bands at the $\Gamma$ point.[15,22,23] Superconductivity occurs for $y > 0.2$ and thus it is found that topological surface states are in proximity to superconducting pairing in the same material, making the surface state of Fe(Te,Se) a candidate topological superconductor.[22,24,25] Several experiments, using different techniques, have observed signatures of topological superconductivity and MBSs in the Fe(Te,Se) system. Angle-resolved photoemission spectroscopy (ARPES) has been used to observe the Dirac electronic structure with an $s$-wave gap and spin-momentum locking necessary for a topological superconducting state.[15] Scanning tunneling spectroscopy measurements have found zero bias peaks in magnetic field induced vortex cores, indicative of MBS within the vortex.[14,16,17] While the combination of experiments yields strong evidence for the existence of MBS in the Fe(Te,Se) system, experimental artifacts are found that raises questions as to the origins of these signatures. While theory predicts that a vortex in a topological superconductor should create an MBS, only a small fraction of the vortices in Fe(Te,Se) show an isolated, pure MBS.[16] Inhomogeneities in the chemical potential arising from heavy Se doping can create mixed phases and even induce zero energy Andreev bound states that can appear as MBS and raise questions as to the origins of the observed MBS signatures in tunneling experiments.[20,26] Thus, we need a material platform that has a large $T_c$ and large superconducting gap while minimizing inhomogeneities from chemical doping to isolate and interrogate pure MBS for potential applications.

The recent discovery of the ability to synthesize high quality epitaxial FeTe$_{1-y}$Se$_y$ films on Bi$_2$Te$_3$ offers a new avenue to explore topological superconductivity.[27] While the Fe(Te,Se) has a 4-fold interface compared to the topological insulator's 6-fold interface, the lattice constants of the two materials fortuitously match to generate epitaxial heterojunctions with Fe(Te,Se) $T_c$ ~13 K at optimal doping.[27-29] Interestingly, the heterostructure remains superconducting with $T_c > 10$ K at very low Se ($y = 0.03$) doping.[27] This raises the possibility of creating a topological superconducting system with $T_c \sim 10$ K and significantly less disorder from chemical inhomogeneities. The reduction in disorder could improve the interrogation MBS and their use in quantum computing and sensing applications. Here we use spin resolved ARPES (SARPES) to investigate the spin and electronic structure of epitaxial Fe(Te,Se) thin films grown on Bi$_2$Te$_3$. We find a bulk-like electronic structure of Fe(Te,Se) films as well as the topological interfacial Bi$_2$Te$_3$ Dirac states at the Fermi surface. Surprisingly, we find the Fe(Te,Se) films are superconducting down to the single monolayer (ML) limit. However, the momentum space proximity of the Fe(Te,Se) and topological states can overlap and disrupt the spin texture in the topological states measured by SARPES. Nonetheless, the electronic structure of both Bi$_2$Te$_3$ and Fe(Te,Se) are tunable and an idealized electronic structure is found that shows clear spin-momentum locking in a superconducting heterostructure with significantly lower Se doping than the bulk Fe(Te,Se) counterpart.

**Results**

High quality Fe(Te,Se) overlayers of different thicknesses and Se concentrations were synthesized on both as-grown and annealed Bi$_2$Te$_3$ films grown on Al$_2$O$_3$ substrates, as shown in

Figure 1. The growth was performed using molecular beam epitaxy and subsequently transferred *in-situ* to the SARPES chamber as detailed in the Methods section. The Fe(Te,Se) and $Bi_2Te_3$ electronic structures are quite different as shown schematically in Figure 1a. However, both electronic structures are tunable based on growth conditions. The electronic structure of the as-grown $Bi_2Te_3$ films shown in Figure 1b and 1c are similar to previous results but the Dirac point is observed to be $E - E_F \sim -0.15$ eV and there is no evidence of the bulk conduction bands typically observed from cleaved bulk crystals due to Te vacancies.[30] However, the Fermi level for the as-grown films is located within the valence bands as indicated by the 6-fold symmetric lobes observed in Figure 1b. The bands at the Fermi energy can be systematically tuned by annealing the $Bi_2Te_3$ film after growth in vacuum without Te flux. The resulting creation of Te vacancies dopes the film and raises the Fermi level with respect to the band structure. As shown in Figure 1d and 1e, annealing the $Bi_2Te_3$ film for 30 minutes at $T = 215$ C moves the Dirac point down to $E - E_F \sim -0.5$ eV and the Fermi level crosses through the bulk conduction bands. Thus, post growth annealing can be used as a control parameter for the Fermi crossing of the topological surface states.

For the thicker ~20 ML Fe(Te,Se) films grown on $Bi_2Te_3$, the bulk-like Fe(Te,Se) electronic structure was observed with no evidence of the topological interface states from $Bi_2Te_3$ as shown in Figure 1f-i.[13,15,31] The lack of topological interface states is due to the limited probing depth of the ARPES technique.[32] The bulk electronic structure at the Brillouin zone center $\Gamma$ point consists of three concentric hole pockets from the $d_{xz}$, $d_{yz}$ and $d_{xy}$ bands, similar to previous reports on bulk Fe(Te,Se) samples.[13,15] For the 20 ML Fe(Te,Se) films with $y = 0.25$ shown in Figure 1f and 1g, only the $d_{xy}$ band is observed crossing the Fermi level with the $d_{xz}$ and $d_{yz}$ band maximums within 20 meV of $E_F$. Intensity for the $d_{xy}$ band is only observed at the Fermi level, see also Figure 4b, which is consistent with previous reports.[13] The circular Fermi surface consists of lobes of intensity centered along $k_y = 0$ that quickly decrease away from $k_y = 0$ due to photoemission matrix elements and the use of linearly polarized light.

For the $y = 0.1$ electronic structure shown in Figure 1h and 1i, there are several differences that should be noted when compared to the $y = 0.25$ case. The $d_{yz}$ band remains relatively unchanged but the $d_{xy}$ band is no longer visible. There is also significant intensity at the Fermi energy at the

$\Gamma$ point that was not present in $y = 0.25$. Previous reports of the electronic structure of bulk Fe(Te,Se) shows broad, incoherent spectral weight centered at $\Gamma$ for $y = 0$ that becomes sharper as $y$ increases.[31] For our $y = 0.1$ films, we can see that the $d_{xz}$ band moves up in energy and touches the Fermi level. Hence both the $y = 0.1$ and $y = 0.25$ Fermi surfaces consist of hole pockets centered at $\Gamma$ but from different bands. The $y = 0.25$ Fermi surface is due to the $d_{xy}$ band while the smaller $y = 0.1$ Fermi surface is due to the $d_{xz}$ band with a band maximum at the Fermi energy. The constant energy maps for $y = 0.25$ and $y = 0.1$ shown in Figures 1f and 1h respectively, reveal concentric circular hole pockets arising from the $d_{xz}$ and $d_{yz}$ bands. The size of the pockets from the $d_{yz}$ band is similar for both $y$ but the pocket from the $d_{xz}$ band is much smaller for $y = 0.1$.

To understand how the topological interfacial states align with the bands from the superconducting overlayer, Fe(Te,Se) films with thickness slightly over a monolayer (nominally 1.25 ML as measured with a quartz crystal monitor) were grown on the annealed $Bi_2Te_3$ films. This thickness was chosen to ensure complete coverage of the $Bi_2Te_3$ surface while limiting the thickness to observe clear signatures of the interfacial topological states as shown in Figure 2. The interfacial electronic structure for $y = 0.25$ is shown in Figure 2b. A clear superposition of the Fe(Te,Se) and $Bi_2Te_3$ electronic structures are observed revealing the $d_{xz}$, $d_{yz}$ and $d_{xy}$ Fe(Te,Se) bands along with the bulk and topological interfacial states from $Bi_2Te_3$. There is additional structure in the spectra below $E - E_F \sim -0.3$ eV away from the $\Gamma$ point that is observed for the 1.25 ML Fe(Te,Se) sample in Figure 2b but not for the 20 ML FTS sample with the same $y$ in Figure 2a. These additional bands are consistent with the bulk $Bi_2Te_3$ valence bands.[30] The $Bi_2Te_3$ were annealed in vacuum prior to growth of the Fe(Te,Se) and the Dirac point is located at $E - E_F \sim -0.35$ eV. However, an additional set of dispersing bands is also observed around $E - E_F \sim -0.5$ eV but is much weaker in intensity. These additional states are likely arising from quantum confinement due to the ultra-thin Fe(Te,Se) overlayer as has been observed previously for $Bi_2Te_3$ exposed to $N_2$ or atmosphere.[33]

For the Fe(Te,Se)/$Bi_2Te_3$ interface with $y = 0.1$ shown in Figure 2d, the linear dispersion from the $Bi_2Te_3$ topological interface states are also observed. The $Bi_2Te_3$ annealing conditions were the same for the two cases and the Dirac point is similarly located at $E - E_F \sim -0.35$ eV.

However, the clarity of the linearly dispersing bands and the replica bands from quantum confinement are far superior for the $y = 0.1$ case. It is unclear if there is Se intermixing with the Te in the $Bi_2Te_3$ during Fe(Te,Se) growth, but the improved clarity of the topological bands is indicative of a sharper interface with less interfacial disorder.

To realize proximity induced topological superconductivity, the spin-momentum locking in the topological Dirac states must exist at the interface. To interrogate the spin of the electrons, SARPES measurements are performed at a fixed energy while scanning across different momentum. To benchmark the expected spin texture for our $Bi_2Te_3$ films, SARPES measurements were performed on an annealed $Bi_2Te_3$ film without any Fe(Te,Se) overlayer as shown in Figure 3. The momentum distribution for the spin resolved intensity is shown in Figure 3b where $I_y^+$ and $I_y^-$ indicates the electron spin intensity along the into and out-of page directions, respectively, as shown in Figure 3a. The spin polarization from the spin intensities is determined by the asymmetry in the spin signal using $P = 1/S_{eff}(\frac{I_y^+ - I_y^-}{I_y^+ + I_y^-})$ where $S_{eff} = 0.275$ is the effective Sherman function for our ferrum spin detector.[34-37] For the annealed $Bi_2Te_3$ sample, we find a ~20% spin polarization along the $k_y$ direction on each side of the topological interface states as shown in Figure 3c. While this is significantly lower than the theoretical 100% spin polarization, it is similar to previous measurements on $Bi_2Te_3$ samples and is in part due to the hexagonal warping of the topological states for $Bi_2Te_3$.[38]

The transport and SARPES measurements for $y = 0.25$ and 0.1 are shown in Figures 4a and 5a respectively. Remarkably, superconductivity is observed for both the 1.25 ML Fe(Te,Se) films grown on $Bi_2Te_3$. For $y = 0.25$, the downward turn in resistivity indicative of the onset of superconductivity occurs at $T_c^{onset} \sim 11$ K while y = 0.1 reveals $T_c^{onset} \sim 8$ K. However, both samples show zero resistance at $T_c^{R=0} \sim 0.5$ K and a reduction of $T_c^{R=0}$ under magnetic field as expected for superconductivity. The existence of monolayer superconductivity is surprising and while this study focuses on monolayer systems to reveal the interfacial electronic structure, previous reports demonstrate thicker films have higher $T_c$s for similar $y$ as well as sharper superconducting phase transitions.[27] It should be noted that superconductivity is measured *ex situ* after capping the heterostructure with an amorphous Te overlayer. Thus, the monolayer superconducting atomic and electronic structure is robust. Monolayer FeSe grown on $SrTiO_3$, another thin film superconductor in the iron pnictide family, typically shows reduced $T_c$ when compared to the superconducting gap opening temperature and a similarly broad transition when measured by transport.[39-42] It is not clear if the broad transition is due to the reduced dimensionality and/or the capping layer required for *ex situ* transport measurements. In addition, it should be emphasized that both $y = 0.25$ and $y = 0.1$ monolayer Fe(Te,Se) films show superconductivity despite the differences in the electronic bands crossing the Fermi level.

The Fermi surface for the Fe(Te,Se)/$Bi_2Te_3$ films with $y = 0.25$ and 0.1, shown in Figure 4c and Figure 5c respectively, reveal the coexistence of a circular band from the Fe(Te,Se) film as well as the hexagonally warped topological interface states from the underlying $Bi_2Te_3$. The primary difference is the relative position in momentum for the Fe(Te,Se) and $Bi_2Te_3$ bands crossing the Fermi energy. For $y = 0.25$ Fermi surface shown in Figure 4c, the size of the $d_{xy}$ hole pocket is similar to that observed in the thicker 20ML Fe(Te,Se) films shown in Figure 1f. However, the hexagonal Fermi surface from the underlying $Bi_2Te_3$ Dirac states overlaps with the Fe(Te,Se) $d_{xy}$ band. This is emphasized in the dispersion for $y = 0.25$ shown in Figure 4d, where the Fe(Te,Se) and topological surface states cross at essentially the same Fermi wavevector $k_F$. For the $y = 0.1$ Fermi surface shown in Figure 5c, the disappearance of the $d_{xy}$ band as well as the concentration of intensity centered at $\Gamma$ due to the $d_{xz}$ band makes the circular Fe(Te,Se) hole pocket much smaller and well within the hexagonal topological surface states. In the dispersion shown in Figure 5d, the Fe(Te,Se) and topological surface states have distinct $k_F$s.

To form a topological superconducting state in a hybrid *s*-wave superconductor and topological insulator interface that can host MBS, cooper pairs formed by electrons at $+k$ and $-k$ in momentum space must have a singly degenerate helical spin texture.[4,12] To understand the potential for the Fe(Te,Se)/$Bi_2Te_3$ heterojunction to host MBS, SARPES measurements were conducted at the Fermi level. Figure 4 and Figure 5 show the SARPES measurements for $y = 0.25$ and $y = 0.1$, respectively. For $y = 0.25$ spin intensity in Figure 4e, we are unable to resolve any asymmetry in the $I_y^+$ and $I_y^-$ intensities resulting in negligible spin polarization shown in Figure 4f. The same SARPES measurement was performed on the $y = 0.1$ sample as shown in Figure 5e and 5f. In this case, an asymmetry is observed in $I_y^+$ and $I_y^-$ which results in a clear spin polarization, of similar magnitude as the $Bi_2Te_3$ reference sample in Figure 3c. Due to the momentum resolution of the spin intensity measurements, we are unable to resolve the individual Fe(Te,Se) and $Bi_2Te_3$ bands in Figure 5e. Nonetheless, a clear spin-momentum locking is evident in the $y$ = 0.1 sample revealed by the SARPES data.

**Discussion**

Bulk FeSe is a superconductor and the addition of Te simultaneously introduces a large spin orbit coupling and pushes the $p_z$ band downward in energy towards $E_F$ which creates a band inversion and nontrivial topology for the system.[15,22,23,25,43] While non-trivial topology is theoretically predicted for $y$ < 0.7,[23] the interplay between superconductivity and topology is much more nuanced as revealed by numerous coexisting superconducting, non-superconducting and topological phases arising from chemical inhomogeneities in doped samples.[20] These inhomogeneities result in inconsistent MBS observations within magnetic field induced vortex cores and raises questions as to the existence of MBS for the system.[16,26] For our Fe(Te,Se)/$Bi_2Te_3$ system, we do not observe topological surface states on the surfaces of our 20 ML Fe(Te,Se) films. However, we do find superconductivity down to the ML limit.

The topological Dirac electronic states on the boundary of a topological insulator arise due to the interface between topologically nontrivial and topological trivial systems. The heterojunction between Fe(Te,Se) and $Bi_2Te_3$ creates the necessary boundary conditions to realize the topological Dirac band structure at this buried interface. Hence, for topological superconductivity the needed spin-momentum locked topological band structure is provided by the Fe(Te,Se)/$Bi_2Te_3$ interface which is directly adjacent to the superconducting Fe(Te,Se) overlayer. Previous results show superconductivity with $T_c$ ~ 10 K in Fe(Te,Se)/$Bi_2Te_3$ down to $y$ ~ 0.03 and our 1.25 ML Fe(Te,Se) films with $y$ = 0.1 on $Bi_2Te_3$ are still superconducting.[27] With reduced $y$ and nonrelevance of the topological states in the Fe(Te,Se) film itself, an idealized Fu-Kane hybrid proximity induced topological superconductor can be realized while minimizing inhomogeneities from doping.[12]

For a hybrid proximity induced topological superconductor spin non-degenerate states are needed to form cooper pairs in the Dirac cone to generate *p*-wave superconducting pairing.[4,12] Thus, the Cooper pairs formed by coupling electrons at $+k$ and $-k$ must have the necessary nondegenerate spin state. For the Fe(Te,Se) films with $y = 0.25$, the overlap of the spin degenerate Fe(Te,Se) states with the topological states from the underlying $Bi_2Te_3$ disrupts the unique topological spin texture at $+k$ and $-k$. However, by tuning the electronic structure of both Fe(Te,Se) and $Bi_2Te_3$, the necessary spin-momentum locking at $+k$ and $-k$ can be restored while still in proximity to a superconductor.

Beyond the tunability of the superconducting and topological electronic structure, this system has other advantages over other proposed hybrid topological superconducting systems. Three dimensional topological insulating materials need to be thicker than a critical thickness to prevent hybridization of the topological electronic structure at the top and bottom boundaries of the material.[44,45] However, it has also been shown that the proximity induced superconductivity pairing is quickly suppressed with thickness.[46] Hence, many approaches of coupling a topological insulator to an *s*-wave superconductor and searching for topological superconductivity on the surface of the topological insulator faces challenges due to these competing thickness requirements for the hybrid system. For the Fe(Te,Se)/$Bi_2Te_3$ system, we probe the interface from the side of the superconductor and thicker $Bi_2Te_3$ films can be used which prevents overlap with the topological states at the $Bi_2Te_3$/substrate interface. In addition, the topological states are directly adjacent to the superconducting wavefunction where proximity effects are maximum. Since magnetic field induced vortices in a superconducting film penetrate the entire film, the MBS from the topological superconducting state should be accessible at this buried interface.

We observe that the monolayer Fe(Te,Se)/$Bi_2Te_3$ has the necessary ingredients for topological superconductivity. As a next step, investigations to observe the zero bias MBS peaks in the vortex cores using scanning tunneling spectroscopy or non-local transport methods are required.[47] In addition, the relative energy scales of the superconducting gap and Fermi level of the topological states are also important for optimizing the topological superconducting state.[48] Hence, further studies of MBS in the Fe(Te,Se)/$Bi_2Te_3$ system are needed to determine the optimum electronic structure for realizing robust MBS in this system. However, we have demonstrated the ability to tune both the Fe(Te,Se) and Bi$_2$Te$_3$ electronic structures while simultaneously minimizing inhomogeneities from doping in the Fe(Te,Se) films.

**Conclusions**

We have investigated the spin resolved electronic structure of Fe(Te,Se) films grown on Bi$_2$Te$_3$. We observe the Fe(Te,Se) is superconducting down to the monolayer limit but $T_c^{R=0}$ is lower than observed for thicker films. This monolayer superconductivity is robust enough for *ex situ* measurements and is observed for $y = 0.25$ and $y = 0.1$ thin films with different Fe(Te,Se) Fermi surfaces. For the thin 1.25 ML Fe(Te,Se) films we observe the Bi$_2$Te$_3$ topological interface states. For Fe(Te,Se) films with $y = 0.25$ we find overlapping electronic structures of the Fe(Te,Se) and Bi$_2$Te$_3$ which disrupts the spin-momentum locking observed in the topological states by SARPES. However, the electronic structure of both Fe(Te,Se) and Bi$_2$Te$_3$ can be tuned and optimized to reveal the needed spin-momentum locking. We have demonstrated Fe(Te,Se)/Bi$_2$Te$_3$ is a flexible system that can support MBS with expected $T_c$s above that of *s*-wave superconductors like Nb while simultaneously minimizing chemical inhomogeneities from Se doping.

**Methods**

*Thin Film Synthesis and Transport:* Films were grown by molecular beam epitaxy on c-plane sapphire following a similar recipe previously reported.[27] The substrates were mounted on holders and cleaned in UV ozone prior to being pumped down and then transferred into the MBE chamber. They were then heated to 600 °C in Te for 15 minutes, and subsequently cooled to 145 °C. The Bi$_2$Te$_3$ was grown using a two-step growth processes where an initial seed layer of 3 quintuple layers (QL, 1 QL ≈ 1 nm) of Bi$_2$Te$_3$ was deposited and subsequently heated to 225 °C where the remaining 9 QL were deposited.[49,50] The samples were idled at 225 °C for 20 minutes with in a Te flux, then the 5 minutes in no Te flux. This was followed by cooling to 200 °C with no Te flux, at which point the Fe(Te,Se) was grown with the Te:Se ratio set to give the correct stoichiometry. Once the Fe(Te,Se) growth finished, the Te and Se were closed and the films were cooled. Transport measurements were performed using pressed indium contacts in van der Pauw geometry. The data down to 2 K was obtained using a Quantum Design Physical Property Measurement System with a standard AC bridge. The data from 1.5 K down to 0.02 K was measured in an Oxford Triton dilution refrigerator using a lock-in amplifier setup.

*Spin and Electronic Structure Measurements:* Spin and angle resolved photoemission spectroscopy measurements were performed in a lab-based system, coupled to the molecular beam epitaxy system, using a Scienta DA30L hemispherical analyzer with a base pressure of $P = 5 \times 10^{-11}$ Torr and a base temperature of T ~ 7 K. Samples were illuminated with linearly polarized light using an Oxide $h\nu = 11$ eV laser system. For electronic dispersion measurements, a pass energy of 5 eV and 0.3 mm slit was used for a total energy resolution ~ 4 meV and momentum resolution ~ 0.01 Å$^{-1}$. Dual VLEED ferrums that utilize exchange scattering are coupled to the electron analyzer and used to determine the spin of the measured electrons. For spin resolved measurements a pass energy of 5 eV and a 1 mm x 3mm spin aperture was used yielding a total energy resolution ~ 35 meV and momentum resolution ~ 0.033 Å$^{-1}$.

## Acknowledgements

We thank Michael McGuire for fruitful discussions. This material is based on work supported by the U.S. Department of Energy, Office of Science, National Quantum Information Sciences Research Centers, Quantum Science Center. H. M. acknowledges support from U.S. DOE, Office of Science, Basic Energy Sciences, Materials Science and Engineering Division. The work at Rutgers is supported by National Science Foundation's DMR2004125, Army Research Office's W911NF2010108, and the center for Quantum Materials Synthesis (cQMS), funded by the Gordon and Betty Moore Foundation's EPiQS initiative through grant GBMF10104.

## References

[1]   A. Y. Kitaev, *Ann. Phys.* **2003**, *303*, 2.
[2]   C. Nayak, S. H. Simon, A. Stern, M. Freedman, and S. Das Sarma, *Rev. Mod. Phys.* **2008**, *80*, 1083.
[3]   N. Read and D. Green, *Phys. Rev. B* **2000**, *61*, 10267.
[4]   M. Sato and Y. Ando, *Rep. Prog. Phys.* **2017**, *80*, 076501.
[5]   P. W. Anderson, *J. Phys. Chem. Solid* **1959**, *11*, 26.
[6]   Y. S. Hor, A. J. Williams, J. G. Checkelsky, P. Roushan, J. Seo, Q. Xu, H. W. Zandbergen, A. Yazdani, N. P. Ong, and R. J. Cava, *Phys. Rev. Lett.* **2010**, *104*, 057001.

[7] L. A. Wray, S.-Y. Xu, Y. Xia, Y. S. Hor, D. Qian, A. V. Fedorov, H. Lin, A. Bansil, R. J. Cava, and M. Z. Hasan, *Nat. Phys.* **2010**, *6*, 855.
[8] L. Fu and E. Berg, *Phys. Rev. Lett.* **2010**, *105*, 097001.
[9] S. Ran, C. Eckberg, Q. P. Ding, Y. Furukawa, T. Metz, S. R. Saha, I. L. Liu, M. Zic, H. Kim, J. Paglione, and N. P. Butch, *Science* **2019**, *365*, 684.
[10] L. Andersen, A. Ramires, Z. Wang, T. Lorenz, and Y. Ando, *Sci. Adv.* **2020**, *6*, eaay6502.
[11] P. F. S. Rosa, A. Weiland, S. S. Fender, B. L. Scott, F. Ronning, J. D. Thompson, E. D. Bauer, and S. M. Thomas, *Commun. Mater.* **2022**, *3*.
[12] L. Fu and C. L. Kane, *Phys. Rev. Lett.* **2008**, *100*, 096407.
[13] H. Miao, P. Richard, Y. Tanaka, K. Nakayama, T. Qian, K. Umezawa, T. Sato, Y. M. Xu, Y. B. Shi, N. Xu, X. P. Wang, P. Zhang, H. B. Yang, Z. J. Xu, J. S. Wen, G. D. Gu, X. Dai, J. P. Hu, T. Takahashi, and H. Ding, *Phys. Rev. B* **2012**, *85*.
[14] J. X. Yin, Z. Wu, J. H. Wang, Z. Y. Ye, J. Gong, X. Y. Hou, L. Shan, A. Li, X. J. Liang, X. X. Wu, J. Li, C. S. Ting, Z. Q. Wang, J. P. Hu, P. H. Hor, H. Ding, and S. H. Pan, *Nat. Phys.* **2015**, *11*, 543.
[15] P. Zhang, K. Yaji, T. Hashimoto, Y. Ota, T. Kondo, K. Okazaki, Z. Wang, J. Wen, G. D. Gu, H. Ding, and S. Shin, *Science* **2018**, *360*, 182.
[16] D. Wang, L. Kong, P. Fan, H. Chen, S. Zhu, W. Liu, L. Cao, Y. Sun, S. Du, J. Schneeloch, R. Zhong, G. Gu, L. Fu, H. Ding, and H. J. Gao, *Science* **2018**, *362*, 333.
[17] T. Machida, Y. Sun, S. Pyon, S. Takeda, Y. Kohsaka, T. Hanaguri, T. Sasagawa, and T. Tamegai, *Nat. Mater.* **2019**, *18*, 811.
[18] X. L. Peng, Y. Li, X. X. Wu, H. B. Deng, X. Shi, W. H. Fan, M. Li, Y. B. Huang, T. Qian, P. Richard, J. P. Hu, S. H. Pan, H. Q. Mao, Y. J. Sun, and H. Ding, *Phys. Rev. B* **2019**, *100*.
[19] C. Chen, K. Jiang, Y. Zhang, C. Liu, Y. Liu, Z. Wang, and J. Wang, *Nat. Phys.* **2020**, *16*, 536.
[20] Y. Li, N. Zaki, V. O. Garlea, A. T. Savici, D. Fobes, Z. Xu, F. Camino, C. Petrovic, G. Gu, P. D. Johnson, J. M. Tranquada, and I. A. Zaliznyak, *Nat. Mater.* **2021**, *20*, 1221.
[21] X. Wu, X. Liu, R. Thomale, and C. X. Liu, *Natl. Sci. Rev.* **2022**, *9*, nwab087.
[22] Z. Wang, P. Zhang, G. Xu, L. K. Zeng, H. Miao, X. Xu, T. Qian, H. Weng, P. Richard, A. V. Fedorov, H. Ding, X. Dai, and Z. Fang, *Phys. Rev. B* **2015**, *92*.
[23] X. Wu, S. Qin, Y. Liang, H. Fan, and J. Hu, *Phys. Rev. B* **2016**, *93*.
[24] N. Katayama, S. Ji, D. Louca, S. Lee, M. Fujita, T. J. Sato, J. Wen, Z. Xu, G. Gu, G. Xu, Z. Lin, M. Enoki, S. Chang, K. Yamada, and J. M. Tranquada, *J. Phys. Soc. Japan* **2010**, *79*.
[25] G. Xu, B. Lian, P. Tang, X. L. Qi, and S. C. Zhang, *Phys. Rev. Lett.* **2016**, *117*, 047001.
[26] Z. Hou and J. Klinovaja, *arXiv:2109.08200* **2021**.
[27] X. Yao, M. Brahlek, H. T. Yi, D. Jain, A. R. Mazza, M. G. Han, and S. Oh, *Nano Lett.* **2021**, *21*, 6518.
[28] H. Qin, X. Chen, B. Guo, T. Pan, M. Zhang, B. Xu, J. Chen, H. He, J. Mei, W. Chen, F. Ye, and G. Wang, *Nano. Lett.* **2021**, *21*, 1327.
[29] X. Yao, A. R. Mazza, M. G. Han, H. T. Yi, D. Jain, M. Brahlek, and S. Oh, *Nano. Lett.* **2022**.
[30] Y. L. Chen, J. G. Analytis, J. H. Chu, Z. K. Liu, S. K. Mo, X. L. Qi, H. J. Zhang, D. H. Lu, X. Dai, Z. Fang, S. C. Zhang, I. R. Fisher, Z. Hussain, and Z. X. Shen, *Science* **2009**, *325*, 178.
[31] E. Ieki, K. Nakayama, Y. Miyata, T. Sato, H. Miao, N. Xu, X. P. Wang, P. Zhang, T. Qian, P. Richard, Z. J. Xu, J. S. Wen, G. D. Gu, H. Q. Luo, H. H. Wen, H. Ding, and T. Takahashi, *Phys. Rev. B* **2014**, *89*.
[32] A. Damascelli, Z. Hussain, and Z.-X. Shen, *Rev. Mod. Phys.* **2003**, *75*, 473.
[33] C. Chen, S. He, H. Weng, W. Zhang, L. Zhao, H. Liu, X. Jia, D. Mou, S. Liu, J. He, Y. Peng, Y. Feng, Z. Xie, G. Liu, X. Dong, J. Zhang, X. Wang, Q. Peng, Z. Wang, S. Zhang, F. Yang, C. Chen, Z. Xu, X. Dai, Z. Fang, and X. J. Zhou, *Proc. Natl. Acad. Sci. U S A* **2012**, *109*, 3694.
[34] R. Bertacco, D. Onofrio, and F. Ciccacci, *Rev. Sci. Instrum.* **1999**, *70*, 3572.
[35] T. Okuda, Y. Takeichi, Y. Maeda, A. Harasawa, I. Matsuda, T. Kinoshita, and A. Kakizaki, *Rev. Sci. Instrum.* **2008**, *79*, 123117.

[36]  A. Winkelmann, D. Hartung, H. Engelhard, C. T. Chiang, and J. Kirschner, *Rev. Sci. Instrum.* **2008**, *79*, 083303.
[37]  M. Escher, N. B. Weber, M. Merkel, L. Plucinski, and C. M. Schneider, *e-J. Surf. Sci. Nanotech.* **2011**, *9*, 340.
[38]  D. Hsieh, Y. Xia, D. Qian, L. Wray, J. H. Dil, F. Meier, J. Osterwalder, L. Patthey, J. G. Checkelsky, N. P. Ong, A. V. Fedorov, H. Lin, A. Bansil, D. Grauer, Y. S. Hor, R. J. Cava, and M. Z. Hasan, *Nature* **2009**, *460*, 1101.
[39]  Q.-Y. Wang, Z. Li, W.-H. Zhang, Z.-C. Zhang, J.-S. Zhang, W. Li, H. Ding, Y.-B. Ou, P. Deng, K. Chang, J. Wen, C.-L. Song, K. He, J.-F. Jia, S.-H. Ji, Y.-Y. Wang, L.-L. Wang, X. Chen, X.-C. Ma, and Q.-K. Xue, *Chin. Phys. Lett.* **2012**, *29*.
[40]  J. J. Lee, F. T. Schmitt, R. G. Moore, S. Johnston, Y. T. Cui, W. Li, M. Yi, Z. K. Liu, M. Hashimoto, Y. Zhang, D. H. Lu, T. P. Devereaux, D. H. Lee, and Z. X. Shen, *Nature* **2014**, *515*, 245.
[41]  B. D. Faeth, S. Xie, S. Yang, J. K. Kawasaki, J. N. Nelson, S. Zhang, C. Parzyck, P. Mishra, C. Li, C. Jozwiak, A. Bostwick, E. Rotenberg, D. G. Schlom, and K. M. Shen, *Phys. Rev. Lett.* **2021**, *127*, 016803.
[42]  Y. Sun, W. Zhang, Y. Xing, F. Li, Y. Zhao, Z. Xia, L. Wang, X. Ma, Q. K. Xue, and J. Wang, *Sci. Rep.* **2014**, *4*, 6040.
[43]  P. D. Johnson, H. B. Yang, J. D. Rameau, G. D. Gu, Z. H. Pan, T. Valla, M. Weinert, and A. V. Fedorov, *Phys. Rev. Lett.* **2015**, *114*, 167001.
[44]  Y. Zhang, K. He, C.-Z. Chang, C.-L. Song, L.-L. Wang, X. Chen, J.-F. Jia, Z. Fang, X. Dai, W.-Y. Shan, S.-Q. Shen, Q. Niu, X.-L. Qi, S.-C. Zhang, X.-C. Ma, and Q.-K. Xue, *Nat. Phys.* **2010**, *6*, 584.
[45]  M. Neupane, A. Richardella, J. Sanchez-Barriga, S. Xu, N. Alidoust, I. Belopolski, C. Liu, G. Bian, D. Zhang, D. Marchenko, A. Varykhalov, O. Rader, M. Leandersson, T. Balasubramanian, T. R. Chang, H. T. Jeng, S. Basak, H. Lin, A. Bansil, N. Samarth, and M. Z. Hasan, *Nat. Commun.* **2014**, *5*, 3841.
[46]  J. A. Hlevyack, S. Najafzadeh, M. K. Lin, T. Hashimoto, T. Nagashima, A. Tsuzuki, A. Fukushima, C. Bareille, Y. Bai, P. Chen, R. Y. Liu, Y. Li, D. Flototto, J. Avila, J. N. Eckstein, S. Shin, K. Okazaki, and T. C. Chiang, *Phys. Rev. Lett.* **2020**, *124*, 236402.
[47]  B. Sbierski, M. Geier, A.-P. Li, M. Brahlek, R. G. Moore, and J. E. Moore, *Phys. Rev. B* **2022**, *106*.
[48]  Y. Zang, F. Kuster, J. Zhang, D. Liu, B. Pal, H. Deniz, P. Sessi, M. J. Gilbert, and S. S. P. Parkin, *Nano Lett.* **2021**, *21*, 2758.
[49]  N. Bansal, Y. S. Kim, E. Edrey, M. Brahlek, Y. Horibe, K. Iida, M. Tanimura, G.-H. Li, T. Feng, H.-D. Lee, T. Gustafsson, E. Andrei, and S. Oh, *Thin Solid Films* **2011**, *520*, 224.
[50]  M. Brahlek, J. Lapano, and J. S. Lee, *J. App. Phys.* **2020**, *128*.

**Figures**

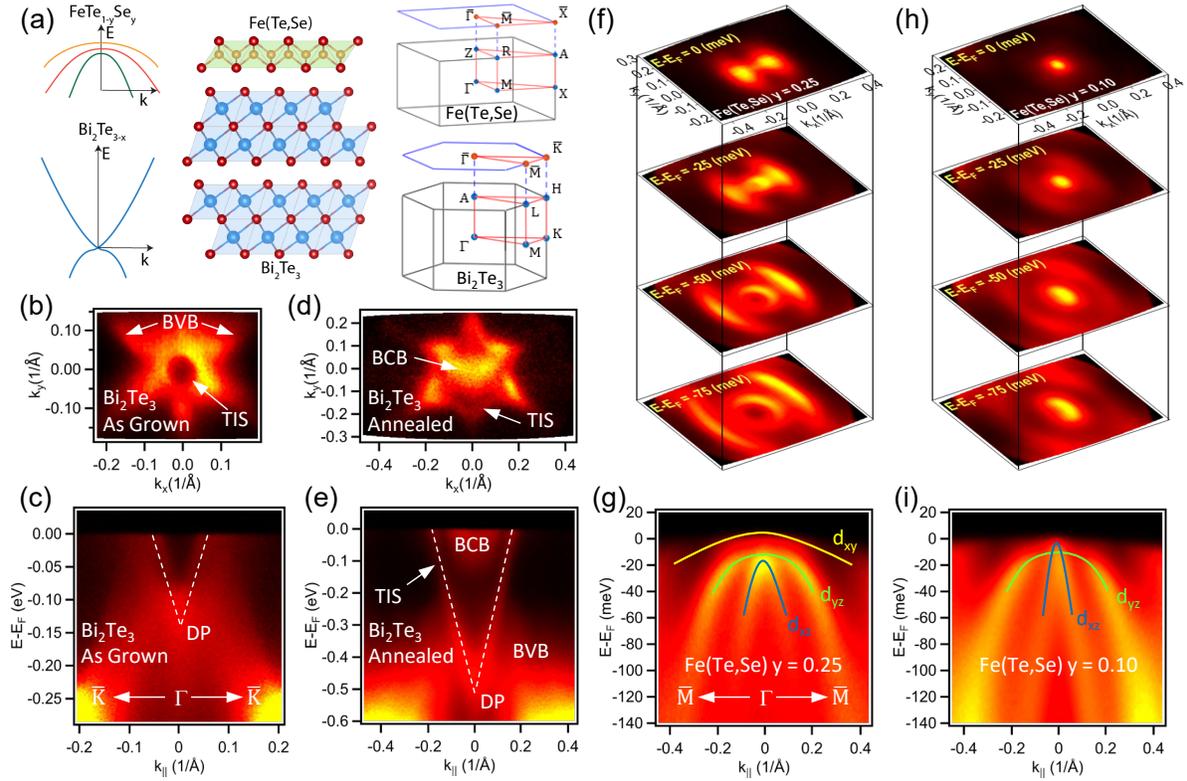

Figure 1: Electronic Structure of Fe(Te,Se) and $Bi_2Te_3$. (a) Schematic representation of the electronic structure (left), atomic structure (middle), and Brillouin zones (right) for Fe(Te,Se)/$Bi_2Te_3$ heterojunction. (b) Fermi surface of the as-grown $Bi_2Te_3$ films with the bulk valence bands (BVB) and topological interface states (TIS) highlighted. (c) Dispersion of the $Bi_2Te_3$ films with the Dirac point (DP) highlighted. The dashed lines are guides to the eye for the TIS. (d) Fermi surface of the $Bi_2Te_3$ film after annealing with the bulk conduction bands (BCB) and TIS highlighted. (e) Dispersion of the $Bi_2Te_3$ films after annealing with the TIS, DP, BCB and BVB highlighted. The dashed lines are guides to the eye for the TIS. (f) Constant energy plots for the 20 ML Fe(Te,Se) films with $y = 0.25$ grown on $Bi_2Te_3$. (g) Dispersion of the 20 ML Fe(Te,Se) films with $y = 0.25$ grown on $Bi_2Te_3$ with guides to the eye for the $d_{xy}$, $d_{yz}$ and $d_{xz}$ bands. (h) Constant energy plots for the 20 ML Fe(Te,Se) films with $y = 0.1$ grown on $Bi_2Te_3$. (i) Dispersion of the 20 ML Fe(Te,Se) films with $y = 0.1$ grown on $Bi_2Te_3$ with guides to the eye for the $d_{yz}$ and $d_{xz}$ bands.

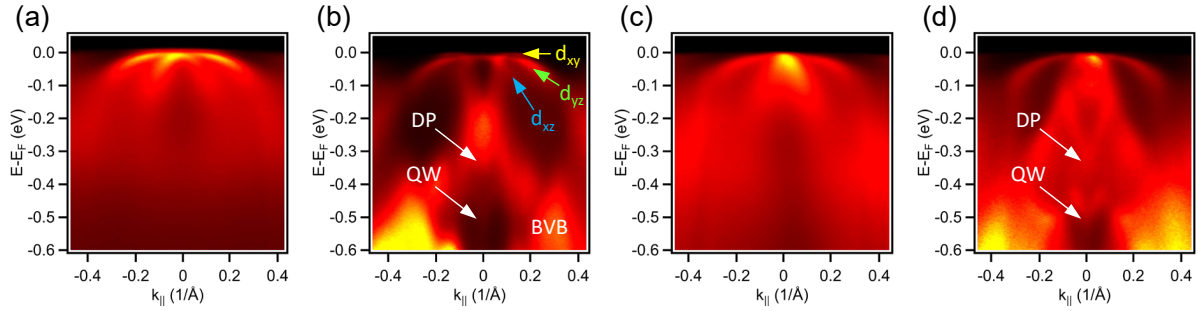

Figure 2: Comparison of 20 ML and 1.25 ML Fe(Te,Se) films grown on $Bi_2Te_3$. (a) Dispersion of 20 ML Fe(Te,Se) films with $y = 0.25$ grown on $Bi_2Te_3$. (b) Dispersion of 1.25 ML Fe(Te,Se) films with $y = 0.25$ grown on $Bi_2Te_3$ with the Fe(Te,Se) $d_{xy}$, $d_{yz}$ and $d_{xz}$ bands highlighted and the $Bi_2Te_3$ BVB, DP and a quantum well (QW) state highlighted. (c) Dispersion of the 20 ML Fe(Te,Se) films with $y = 0.1$ grown on $Bi_2Te_3$. (d) Dispersion of the 1.25 ML Fe(Te,Se) films with $y = 0.1$ grown on $Bi_2Te_3$ with the DP and a QW state highlighted.

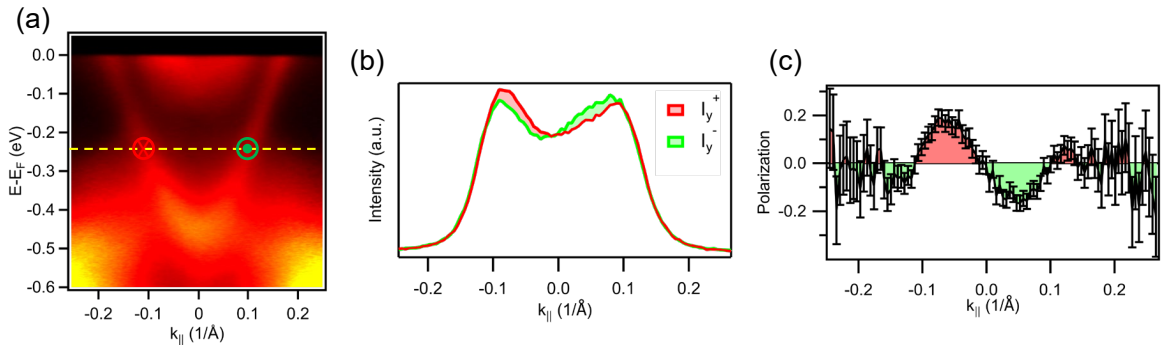

Figure 3: Spin texture of annealed $Bi_2Te_3$ films. (a) Dispersion of the $Bi_2Te_3$ films after annealing. The yellow dashed line indicates the constant energy line used for the momentum dependent spin map in (b) and (c). The red cross and green dot indicate the spin directions into and out-of the page, respectively. (b) Spin dependent scattered intensity for the $I_y^+$ and $I_y^-$ directions. (c) Momentum dependent spin polarization along the constant energy line in (a).

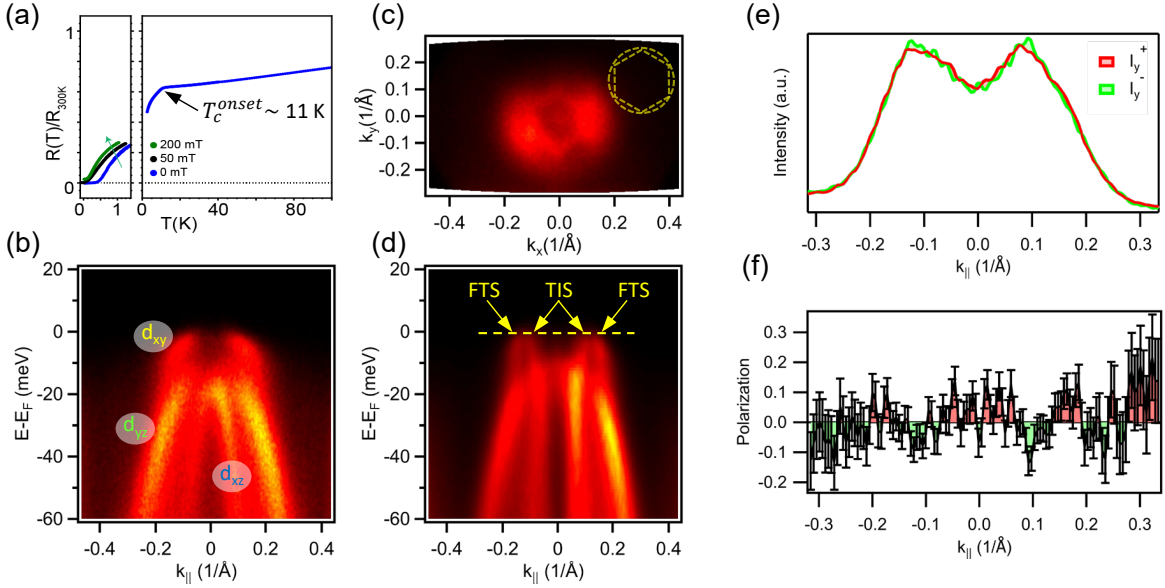

Figure 4: Spin dependent electronic structure for 1.25 ML Fe(Te,Se) films with $y = 0.25$ grown on $Bi_2Te_3$. (a) Transport data for the 1.25 ML Fe(Te,Se) films showing superconducting $T_c^{onset} \sim$ 11 K with zero resistance at T ~ 0.5 K. (b) Dispersion near the Fermi level for 20 ML Fe(Te,Se) films with the $d_{xy}$, $d_{yz}$ and $d_{xz}$ bands annotated. (c) Fermi surface for the 1.25 ML Fe(Te,Se) film grown on $Bi_2Te_3$. The dashed lines are a schematic representation showing the relative positions of the circular Fe(Te,Se) and hexagonal TIS $Bi_2Te_3$ Fermi surfaces found in the center of the figure. (d) Dispersion near the Fermi level for the 1.25 ML Fe(Te,Se) films grown on $Bi_2Te_3$ with the Fe(Te,Se) (FTS) and $Bi_2Te_3$ TIS bands highlighted. The yellow dashed line indicates the constant energy line used for the momentum dependent spin data in (e) and (f). (e) Spin dependent scattered intensity at the Fermi level for the $I_y^+$ and $I_y^-$ directions. (f) Momentum dependent spin polarization along the constant energy line at the Fermi level shown in (d).

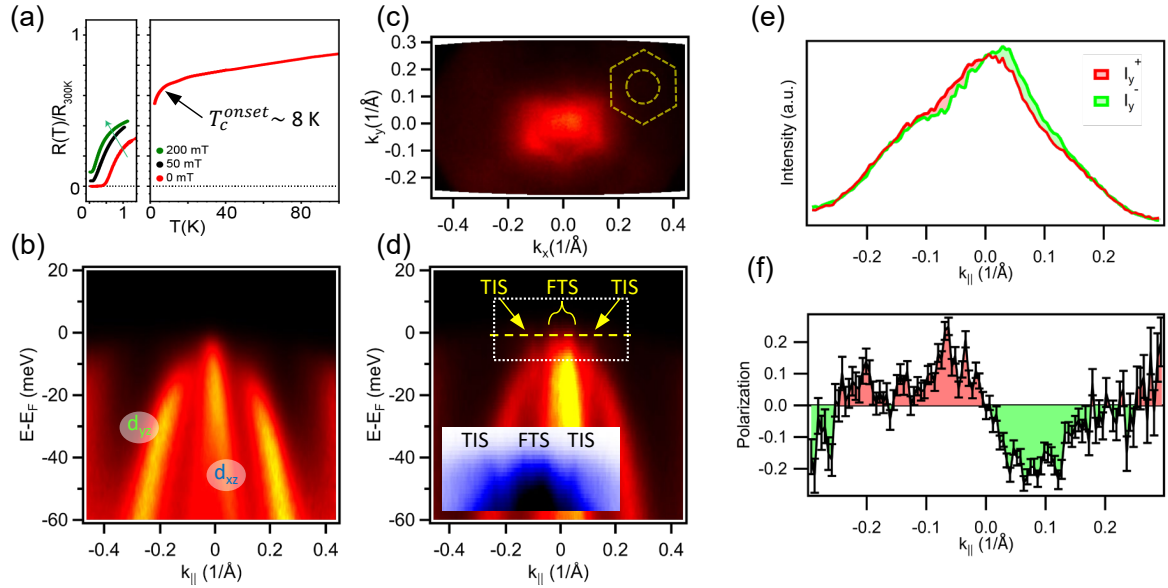

Figure 5: Spin dependent electronic structure for 1.25 ML Fe(Te,Se) films with y = 0.1 grown on $Bi_2Te_3$. (a) Transport data for the 1.25 ML Fe(Te,Se) films showing superconducting $T_c^{onset} \sim 8$ K with zero resistance at T ~ 0.5 K. (b) Dispersion near the Fermi level for 20 ML Fe(Te,Se) films with the $d_{yz}$ and $d_{xz}$ bands annotated. (c) Fermi surface for the 1.25 ML Fe(Te,Se) film grown on $Bi_2Te_3$. The dashed lines are a schematic representation showing the relative positions of the circular Fe(Te,Se) and hexagonal TIS $Bi_2Te_3$ Fermi surfaces found in the center of the figure. (d) Dispersion near the Fermi level for the 1.25 ML Fe(Te,Se) films grown on $Bi_2Te_3$ with the FTS and $Bi_2Te_3$ TIS bands highlighted. The yellow dashed line indicates the constant energy line used for the momentum dependent spin data in (e) and (f). The region at the Fermi level outlined by the white dashed lines is shown in the inset with alternate colormap to highlight the TIS. (e) Spin dependent scattered intensity at the Fermi level for the $I_y^+$ and $I_y^-$ directions. (f) Momentum dependent spin polarization along the constant energy line at the Fermi level shown in (d).